\documentclass[useAMS,usenatbib]{mn2e}
\usepackage{amsfonts}
\usepackage{graphicx,float}
\usepackage[ansinew]{inputenc}
\usepackage{epstopdf}
\usepackage{datetime}
\usepackage{titlesec}
\usepackage{subfig}
\usepackage{caption}
\usepackage[section]{placeins}
\usepackage{color}
\usepackage{multirow}
\usepackage{amsmath}
\usepackage{breqn}
\DeclareMathOperator\arctanh{arctanh}
\onecolumn

\title[Constraining $f(R)$ models]{Constraining $f(R)$ models with cosmic chronometers and the HII galaxy Hubble diagram}
\author[J. Sultana \& M. K. Yennapureddy \& F. Melia \& D. Kazanas]{
Joseph Sultana,$^{1}$\thanks{E-mail: joseph.sultana@um.edu.mt}
Manoj K. Yennapureddy,$^{2}$\thanks{E-mail: manojy@email.arizona.edu}
Fulvio Melia,$^{3}$\thanks{E-mail: fmelia@email.arizona.edu}
Demosthenes Kazanas,$^{4}$\thanks{E-mail: Demos.Kazanas-1@nasa.gov}
\\
$^{1}$Department of Mathematics, Faculty of Science, University of Malta, Msida MSD 2080, Malta\\
$^{2}$Department of Physics, The University of Arizona, Tucson AZ 85721\\
$^{3}$Department of Physics, The Applied Math Program, and Department of Astronomy, The University of Arizona, Tucson AZ 85721\\
$^{4}$Astrophysics Science Division, NASA/Goddard Space Flight Center, Greenbelt, Maryland 20771, USA
}

\begin{document}

\date{}

\pagerange{\pageref{firstpage}--\pageref{lastpage}} \pubyear{2022}

\maketitle


\label{firstpage}

\begin{abstract}

We consider several well-known $f(R)$ cosmological models and constrain their parameters, namely the deviation parameter $b$ and the cosmological parameters $\Omega_m$ and $h$. We first obtain analytical approximations for the Hubble rate $H(z)$ and the luminosity distance $d_L(z)$ in terms of these parameters, and then test these against the observational expansion rate derived from cosmic chronometers and the distance modulus in the HII galaxy Hubble diagram, obtained in a model-independent way using Gaussian Processes (GP). We first optimize the models based solely on the cosmic chronometers and then repeat this process with a joint analysis using both the cosmic chronometers and HII galaxies.

\end{abstract}


\section{Introduction}
Modified gravity theories are a group of alternative theories of gravity, in which the repulsive force producing the accelerated expansion of the Universe in the early and late stages, is a purely geometric effect \citep{clifton12}. These theories thus evade the coincidence and cosmological problems of standard $\Lambda$CDM cosmology in General Relativity (GR) or the origin of some yet to be observed scalar field in the so called scalar-tensor cosmological models \citep{faraoni-book}. In these theories, the simple Lagrangian in the general relativistic Einstein-Hilbert action is replaced with a
nonlinear function $f(R, R_{ab}R^{ab}, R_{abcd}R^{abcd},\cdots)$ of the Ricci scalar $R$, and other curvature invariants, such as $R_{ab}R^{ab}$ and $R_{abcd}R^{abcd}$.

The simplest and most natural extension of GR is $f(R)$-gravity \citep{sotiriou10}, in which the Lagrangian co{\tiny }nsists of some function of the Ricci scalar. This prominent and well studied theory has been used extensively in cosmology, starting with an inflationary model \citep{starobinsky80} of the early Universe having $f(R) = R + \alpha R^2$. An early model with a late time cosmic acceleration was obtained with the choice $f(R) = 1/R$ \citep{carroll04}. This was ruled out rather quickly, however, due to its instabilities \citep{dolgov03,nojiri03} and violation of the post-Newtonian tests of GR \citep{chiba03}. $f(R)$-gravity subsequently became even more popular, with numerous $f(R)$ models being proposed over the past decade. Some of these models have the advantage of describing the entire evolution from the early phase of inflation to the current accelerated expansion \citep{cognola08}. However, many others fail to predict a matter-dominated era \citep{amendola07,amendola07-2} and therefore cannot be interpreted as viable models of gravity. So far there are only a few observationally viable $f(R)$ models in the literature that also satisfy the Solar-system tests. Some popular examples include the Hu-Sawicki model \citep{hu07}, Starobinsky's model \citep{starobinsky07}, Tsujikawa's model \citep{tsujikawa08} and the Exponential model \citep{linder09}.

These models approach $\Lambda$CDM at high redshifts or equivalently large $R$, and differ from the standard model only at low redshifts, $z < 2$. It was originally claimed that these models are independent of $\Lambda$CDM and do not contain a cosmological constant $\Lambda$. It was later shown \citep{bamba13}, however, that these can be represented in a form that differs from $\Lambda$CDM via a so-called deviation parameter $b$, such that $f(R) \rightarrow R - 2\Lambda$, as $b\rightarrow 0$. The free parameters in these models have been well constrained {\citep{basilakos13,nunes17,romero18}} using observations of Type Ia SNe, Baryonic Acoustic Oscillations (BAO), the Cosmic Microwave Background, the Hubble parameter $H(z)$, {and cosmic chronometers (CC) data} and, in some cases \citep{odintsov17,dombriz16}, the fits do not show any statistically significant difference from $\Lambda$CDM.

In this paper, we utilize Gaussian Processes (GP) in a model-independent way to obtain the expansion rate $H(z)$ from cosmic chronometers and the luminosity distance $d_L(z)$ from HII galaxies. These are then compared with the corresponding geometric expressions for some of the above mentioned $f(R)$ models by carrying out a $\chi^2$ minimization in order to obtain the optimized values of the model parameters.

In some of the earlier studies about $f(R)$ models (see \citet{nunes17} for example) the computation of $H(z)$ was based on a mixture of measurements extracted from BAO and cosmic chronometers. The standard procedure
in BAO measurements uses a fiducial cosmological model (taken to be a basic version of $\Lambda$CDM)
in order to transform the observed redshifts and angles to the measured angular and radial BAO peak
positions. Therefore the use of these BAO measurements to obtain $H(z)$ introduces a non-ignorable
model dependence in its calculation. {There are only two or three BAO measurements
that avoid this problem (as described in more detail in \citet{melia-lopez17}) and all these
occur at very low redshifts.} Recently a number of procedures \citep{vargas18,ding18,carter20,heinesen20}
have been suggested to reduce this dependence on the choice of the fiducial cosmological model.
Another advantage of GP is that the errors that are obtained along with the function itself, strike a balance between very smooth and rapidly oscillating variations in the data and this produces a $1\sigma$ confidence region that is significantly tighter than the error that results from, say, a simple $\chi^2$.

In this paper, we obtain the optimized values of the cosmological parameters $\Omega_m$ and $h$ and the deviation parameter $b$ for the two most popular viable $f(R)$ models, namely the Hu-Sawicki and Starobinski models, and another model which is called the ArcTanh model, that was introduced recently in \citet{romero18}. We obtain the functions $H(z)$ and $d_L(z)$ by using Gaussian Processes from the cosmic chronometer and HII galaxy data, and these are then compared with the corresponding analytical expressions for the models to obtain two sets of constraints for the parameters: the first set from the cosmic chronometer data alone; and the second set from a joint analysis of the cosmic chronometer and HII galaxy data. As in most of the earlier studies of these and other $f(R)$ models, we assume spatial flatness, which is a valid assumption considering that these models deviate only slightly from $\Lambda$CDM even at late times. Moreover in our case, including spatial curvature would complicate the fitting procedure for $d_L(z)$, considering that there are three separate analytic expressions for the luminosity distance for the cases of positive, zero and negative spatial curvature.

In the next section, we briefly introduce $f(R)$ gravity and the three $f(R)$ models, and obtain analytic approximations for their expansion rate $H(z)$. Then, in \S~3 we introduce GPs and describe the relevant data sets for cosmic chronometers and HII Galaxies. We present our results in \S~4. In \S~5 we compare the $f(R)$ models with each other and with standard $\Lambda$CDM by computing the Akaike Information Criterion (AIC) and the Bayesian Information Criterion (BIC) from the data. Finally in \S~6 we discuss the results and present our conclusion. Unless otherwise noted, we use units such that $8\pi G = c = 1$.

\section[]{Analytic approximations}

The action of $f(R)$ gravity is given by
\begin{equation}
S = \frac{1}{2}\int\mathrm{d}^4x\sqrt{-g}f(R) + S^{(m)}, \label{fRaction}
\end{equation}
where $f(R)$ is some function of the Ricci scalar $R$ and $S^{(m)}$ is the matter contribution. Varying the action above with respect to the (inverse) metric tensor $g^{ab}$ yields the following field equations
\begin{equation}
f'(R)R_{ab} - \frac{f(R)}{2}g_{ab} = \nabla_{a}\nabla_{b}f'(R) - g_{ab}\Box f'(R) + T_{ab}\,, \label{fReqs}
\end{equation}
where a prime denotes differentiation with respect to $R$, $\Box \equiv g^{ab}\nabla_a\nabla_b$ and $T_{ab}$ is the energy momentum tensor corresponding to the matter part of the action $S^{(m)}$. For the expanding Universe, the source of the energy-momentum tensor is normally taken to be a perfect fluid which includes radiation and cold dark matter, such that $T^{ab} = (p + \rho) u^a u^b + p g^{ab}$, where $u^{a}$ is the four-velocity of the fluid; $u^a u_a = -1$. Also $p = p_m + p_r$ and $\rho = \rho_m + \rho_r$ are the total pressure and energy density of the fluid respectively, expressed in terms of the matter and radiation components, where $p_m=0$ and $p_r = \rho_r/3$. Taking the trace in Equation~(\ref{fReqs}) gives
\begin{equation}
\Box R = \frac{1}{3f''(R)}\left[T - 3f'''(R)(\nabla R)^2 + 2 f(R) - R f'(R)\right], \label{trace}
\end{equation}
with $T \equiv \rho - 3p$ being the trace of $T_{ab}$. Using Equation~(\ref{trace}) in (\ref{fReqs}), the field equations can be expressed in their Einstein form \begin{equation}
G_{ab} = \frac{1}{f'(R)}\left(T_{ab} + T_{ab}^{\mathrm{eff}}\right)\,, \label{einstein-form}
\end{equation}
where
\begin{equation}
T_{ab}^{\mathrm{eff}} = \left[\frac{f(R) - Rf'(R)}{2}g_{ab} + \nabla_{a}\nabla_{b}f'(R)
- g_{ab}\Box f'(R)\right] \label{emcfluid}
\end{equation}
is the energy momentum tensor of the so-called ``curvature fluid'' representing the higher order curvature corrections in the action. From Equation~(\ref{einstein-form}), the effective gravitational
coupling can be taken to be $G_{\mathrm{eff}} \equiv 1/f'(R)$. Also using Equation~(\ref{einstein-form}), it is straightforward to show that $\nabla_a T^{ab}=0$, so that assuming there is no interaction between the matter and radiation components, the Bianchi identity is also valid for the individual components.


The general spatially flat Friedmann-Lema\^itre-Robertson-Walker (FLRW) metric is given by
\begin{equation}
ds^2 = -dt^2 + a(t)^2\left[dr^2 + r^2(d\theta^2 + \sin\theta^2d\phi^2)\right], \label{general-frw}
\end{equation}
where $t$ is the cosmic time and $a(t)$ is the normalized scale factor, such that at the present time $t = t_{0}$, $a(t_0) = 1$.
For this metric, the field Equations~(\ref{einstein-form}) yield the Friedmann equations
\begin{equation}
3H^2 = \frac{1}{f'}\left[\rho + \frac{Rf' - f}{2} - 3Hf''\dot{R}\right], \label{fried}
\end{equation}
and
\begin{equation}
-2f'\dot{H} = \rho_m + (4/3)\rho_r + \ddot{f'} - H\dot{f'} \label{ray}\;,
\end{equation}
where overdot indicates a derivative with respect to cosmic time $t$, such that, e.g., $\dot{R} = \frac{dR}{dt} = a\,H\frac{dR}{da}$. The Ricci scalar can be expressed in terms of the Hubble parameter $H = \dot{a}/{a}$ by
\begin{equation}
R = 6(2H^2 + \dot{H})\,. \label{Ricciscalar}
\end{equation}
Moreover the conservation of the energy momentum for the matter and radiation components leads to
\begin{equation}
\dot{\rho_m} + 3H\rho_m = 0 \quad\quad \dot{\rho_r} + 4H\rho_r = 0 \label{conservation}
\end{equation}

The choice of the function $f(R)$ should be compatible with the evolution of the Universe, as well as the local (solar system) tests. In the literature one can find numerous studies of the confrontation of $f(R)$ gravity with cosmological data. This includes the Cosmic Microwave Background (CMB) data \citep{hwang01}, the Large Scale Structure (LSS) data \citep{abebe13}, Supernova type Ia (SNeIa) \citep{aviles13}, Baryon Acoustic Oscillations (BAO) \citep{santos08}, Cosmic Chronometers (CC) \citep{nunes17}, neutron stars mass-radius data \cite{arapoglu11}, and gravitational lensing data \citep{higuchi16}. Most of these studies use multiple combined data sets. Moreover comparison with solar system data was performed in \citet{ruggiero07,chiba07,amendola08,nojiri07,capozziello08,hu07}. {In a recent paper \citet{negrelli20} the authors use a chameleon-like mechanism to obtain a Post-Newtonian Parameter $\gamma$ for some $f(R)$ models, and show that this is compatible with observational bounds.}
In addition, any $f(R)$ cosmological model must satisfy a number of conditions related to its stability and the prediction of a matter-dominated era (see \citet{hu07,amendola07-2} for details), namely (i) $f'(R) > 0$ for $R \geq R_0 > 0$, where $R_0$ is the Ricci scalar at the present time for the FLRW metric in Equation~(\ref{general-frw}). If the final attractor is de Sitter spacetime with Ricci scalar $R_1$, we also require that $f'(R) > 0$ for $R \geq R_1 > 0$; (ii) $f''(R) > 0$ for $R \geq R_0 > 0$; (iii) $f(R) \approx R - 2\Lambda$ for $R >> R_0$, so that the model reduces to $\Lambda$CDM at early times; and (iv) $0 < \left(\frac{Rf''}{f'}\right)(r) < 1$ at $r = -R f'/f = -2$. Cosmological $f(R)$ models that satisfy the above conditions and observational constraints are considered to be viable models.
\vskip 0.1in
The three viable models we shall constrain with the data are:\\[5mm]
\textit{(i) Hu-Sawicki model.}
The Hu-Sawicki (HS) model \citep{hu07} is given by $f(R) = R - m^2\frac{c_1(R/m^2)^n}{1 + c_2(R/m^2)^n}$, where $c_1$ and $c_2$ are free parameters and $m$ and $n$ are positive constants, with $m^2 \approx \Omega_m^{0}H_0^2$ being of the order of the Ricci scalar $R_0$ at the present time. This can be expressed in $\Lambda$CDM form \citep{bamba13} $f(R) = R - 2\Lambda\left(1 - \frac{1}{1 + (R/b\Lambda)^n}\right)$, with $\Lambda = c_1m^2/2c_2$ and $b=2{c_2^{1 - 1/n}}/{c_1}$. {In our case, we consider the case $n=1$, which is the conventional approach seen in the literature.}\\[5mm]
\textit{(ii) Starobinsky model.}
In the  Starobinsky model \citep{starobinsky07}, $f(R) = R - c_1m^2[1 - (1 + R^2/m^4)^{-n}]$, where $c_1$ is a free parameter and $m$ and $n$ are positive constants, with $m^2 \approx \Omega_m^{0}H_0^2$ as in the HS case. This can also be expressed as a perturbation of $\Lambda$CDM \citep{bamba13}, with $f(R) = R - 2\Lambda\left(1 - \frac{1}{(1 + [\frac{R}{b\Lambda}]^2)^n}\right)$, where $\Lambda = c_1m^2/2$ and $b = 2/c_1$. {Again we take $n=1$.}\\[5mm]
\textit{(iii) ArcTanh model.}
This model was proposed in \citet{romero18} and is given by $f(R) = R - \frac{2\Lambda}{1 + b\arctanh\frac{\Lambda}{R}}$.\\[5mm]
All three models are expressed in terms of a deviation parameter $b$, such that $\lim_{b\rightarrow 0} f(R) = R - 2\Lambda$, which recovers $\Lambda$CDM in the limit of vanishing $b$, while $\lim_{b\rightarrow \infty} f(R) = R$.

In order to obtain analytic approximations for the expansion rate $H(z)$ for the above three models, we follow the procedure outlined in \citet{basilakos13}. We first express the Friedmann equation for a general function $f(R)$ in (\ref{fried}) in terms of $N= \ln a$, where $a$ is the normalized scale factor, such that it takes the form
\begin{equation}
-f_RH^2(N) + (\Omega_{m0} e^{-3N} + \Omega_{r0} e^{-4N})H_0^2 + \frac{1}{6}(f_RR - f) = f_{RR}H^2(N)R'(N), \label{diffeq}
\end{equation}
where prime now indicates differentiation with respect to $N$, $f_R = df/dR$, $f_{RR} = d^2f/dR^2$, and $\Omega_{m0}$, $\Omega_{r0}$ indicate the current values of these cosmological parameters. In this case, the expression for $R(N)$ can be obtained from Equation~(\ref{Ricciscalar}). Since all the above models approach $\Lambda$CDM as $b\rightarrow 0$, one can express the solution to (\ref{diffeq}), $H(N)$, as a Taylor expansion in the deviation parameter, $b$, i.e,
\begin{equation}
H^2(N) = H_{\Lambda}^2(N) + \sum_{i=1}^{M}b^{i}\delta H_{i}^2(N), \label{pert}
\end{equation}
where $H_{\Lambda}(N)$ denotes the Hubble parameter for $\Lambda$CDM universe, given by
\begin{equation}
\frac{H_{\Lambda}^2(N)}{H_0^2} = \Omega_{m0}e^{-3N} + \Omega_{r0}e^{-4N} + (1 - \Omega_{m0} - \Omega_{r0}) \label{LCDMeq}\;,
\end{equation}
and the functions $\delta H_{i}^2(N)$ are obtained by substituting Equation~(\ref{pert}) in (\ref{diffeq}) and obtaining the coefficients of the $b^i$.
It has been shown in \citet{basilakos13} that keeping only the first two terms in the above series expansion (i.e., having $M=2$) would result in an excellent agreement (of better than $0.01\%$; see Fig. 1 in \citet{basilakos13}) between the analytical approximation and the numerical solution over a wide range of redshifts for realistic values of the deviation parameter $b \sim \mathcal{O}(1)$. In our case, we also take $M=2$. For the three $f(R)$ models mentioned above, the expansion in (\ref{pert}) gives the following approximations for the Hubble parameter $H(z)$ (with $N$ re-expressed in terms of the redshift $z$):\\[5mm]
\textit{Hu-Sawicki model}:
\begin{dmath}
\lefteqn{\frac{H(z)^2}{H_0^2}=1-\Omega_m+(1+z)^3 \Omega _m}\\
+\frac{6 b \left(-1+\Omega _m\right)^2 \left(4 \left(-1+\Omega _m\right)^2+(1+z)^3 \left(-1+\Omega _m\right) \Omega _m-2
   (1+z)^6 \Omega _m^2\right)}{(1+z)^9 \left(\frac{4 \left(-1+\Omega _m\right)}{(1+z)^3}-\Omega _m\right)^3}+\frac{b^2 \left(-1+\Omega _m\right)^3}{(1+z)^{24}
   \left(-\frac{4 \left(-1+\Omega _m\right)}{(1+z)^3}+\Omega _m\right)^8} \biggl[1024 \left(-1+\Omega
   _m\right)^6+9216 (1+z)^3 \left(-1+\Omega _m\right)^5 \Omega _m-22848 (1+z)^6 \left(-1+\Omega _m\right)^4 \Omega _m^2+25408 (1+z)^9 \left(-1+\Omega _m\right)^3 \Omega
   _m^3- 7452 (1+z)^{12} \left(-1+\Omega _m\right)^2 \Omega _m^4-4656 (1+z)^{15} \left(-1+\Omega _m\right) \Omega _m^5+37 (1+z)^{18} \Omega _m^6\biggr].\label{approx1}
\end{dmath}
\vspace{5mm}
\textit{Starobinsky model}:
\begin{dmath}
\lefteqn{\frac{H(z)^2}{H_0^2}=1-\Omega _m+(1+z)^3 \Omega _m}\\
+\frac{b^2 \left(-1+\Omega _m\right)^3 \left(32
   \left(-1+\Omega _m\right)^2+32 (1+z)^3 \left(-1+\Omega _m\right) \Omega _m-37 (1+z)^6 \Omega
   _m^2\right)}{(1+z)^{12} \left(-\frac{4 \left(-1+\Omega _m\right)}{(1+z)^3}+\Omega _m\right)^4}+\frac{b^4
   \left(-1+\Omega _m\right)^5}{(1+z)^{30} \left(-\frac{4
   \left(-1+\Omega _m\right)}{(1+z)^3}+\Omega _m\right)^{10}} \biggl[20480 \left(-1+\Omega _m\right)^6+63488 (1+z)^3 \left(-1+\Omega
   _m\right)^5 \Omega _m-234880 (1+z)^6 \left(-1+\Omega _m\right)^4 \Omega _m^2+289024 (1+z)^9
   \left(-1+\Omega _m\right)^3 \Omega _m^3- 44552 (1+z)^{12} \left(-1+\Omega _m\right)^2 \Omega _m^4-82748
   (1+z)^{15} \left(-1+\Omega _m\right) \Omega _m^5+123 (1+z)^{18} \Omega _m^6\biggr].\label{approx2}
\end{dmath}
\vspace{5mm}
\textit{ArcTanh model}:
\begin{dmath}
\lefteqn{\frac{H(z)^2}{H_0^2}=1-\Omega _m+(1+z)^3 \Omega _m}\\
+\frac{2 b \left(-1+\Omega _m\right)^2}{3 (1+z)^{15} \left(\frac{4
   \left(-1+\Omega _m\right)}{(1+z)^3}-\Omega _m\right)^5} \biggl[596 \left(-1+\Omega
   _m\right)^4-109 (1+z)^3 \left(-1+\Omega _m\right)^3 \Omega _m-361 (1+z)^6 \left(-1+\Omega _m\right)^2 \Omega
   _m^2+153 (1+z)^9 \left(-1+\Omega _m\right) \Omega _m^3-18 (1+z)^{12} \Omega _m^4\biggr]+\frac{b^2 \left(-1+\Omega _m\right)^3}{9
   (1+z)^{36} \left(-\frac{4 \left(-1+\Omega _m\right)}{(1+z)^3}+\Omega _m\right)^{12}} \biggl[4189184
   \left(-1+\Omega _m\right)^{10}+23851008 (1+z)^3 \left(-1+\Omega _m\right)^9 \Omega _m-95032704 (1+z)^6 \left(-1+\Omega
   _m\right)^8 \Omega _m^2+149808704 (1+z)^9 \left(-1+\Omega _m\right)^7 \Omega _m^3-110911572 (1+z)^{12} \left(-1+\Omega
   _m\right)^6 \Omega _m^4+28668900 (1+z)^{15} \left(-1+\Omega _m\right)^5 \Omega _m^5+2987537 (1+z)^{18} \left(-1+\Omega
   _m\right)^4 \Omega _m^6-3144240 (1+z)^{21} \left(-1+\Omega _m\right)^3 \Omega _m^7+636102 (1+z)^{24} \left(-1+\Omega
   _m\right)^2 \Omega _m^8-47232 (1+z)^{27} \left(-1+\Omega _m\right) \Omega _m^9+333 (1+z)^{30} \Omega _m^{10}\biggr].\label{approx3}
\end{dmath}
In the above expressions, we have taken $\Omega_{r0} = 0$ in order to simplify the equations. It is clear that these reduce to the corresponding expression for $\Lambda$CDM in (\ref{LCDMeq}) when $b\rightarrow 0$. Having these analytical approximations for $H(z)$ would avoid having to carry out a numerical integration of Equation~(\ref{fried}), which may not always be possible with standard methods, considering that at high redshift, the ODE can become stiff.
For all models, the luminosity distance is then obtained from
\begin{equation}
d_L(z) = \frac{(1+z)}{H_0}\int_0^z\frac{dx}{E(x)}, \label{lumdist}
\end{equation}
where $E(x) = H(x)/H_0$.
\section{Data Analysis}
{\subsection{Gaussian Processes}}
{The Gaussian Process (hereafter GP) reconstructs a function based on the available data. This process doesn't depend on any assumed parametric form, or any model of the data. The  function $f(x)$ for the data at a location $x$ relies on the entire dataset through the underlying kernel or covariance function used in the reconstruction. So for example, the values of the function $f$ at $x$ and $x^{'}$ are related through a kernel as follows
	
\begin{equation}
K(x,x^{'}) = \sigma^{2}_{f}\exp\bigg(-\frac{(x-x^{'})^{2}}{2\delta^{2}}\bigg),
\end{equation}
where $\delta$ and $\sigma^{2}_{f}$ are the hyper-parameters of the kernel, and these are optimized for the given data.  The hyper-parameters $\delta$ and $\sigma^{2}_{f}$  represents the distance over which a significant change in the $x$-direction and $y$-direction occurs respectively. Hence, the kernel depends on the entire dataset and is responsible for smoothing the reconstructed function. Therefore the Gaussian Process (for a detailed description see \citet{seikel12}) is a fully Bayesian, non-parametric technique (doesn't require any parametric model) for reconstructing a function from given data, and avoids the need to use a pre-determined parametric form of the function indicated by a particular model, that may or may not be the correct representation of the said data. In our case this technique leads to a model-independent determination of $H(z)$ and $d_L(z)$ from the cosmic chronometer and HII galaxy observations respectively. The Gaussian Process estimates the new data using the real data. The other significant advantage for using GP, is that it even smooths out the uncertainties in the data. The uncertainties in the real data are propagated into the uncertainties of the GP estimated data points, but these uncertainties depend on the distribution of the data from which the estimates are drawn. Moreover, these 1$\sigma$ confidence regions are very tight and smooth when compared with the rapidly oscillating variations in the actual data. The GP estimated uncertainties along with the GP estimated data points are then used to optimize (maximize) the likelihood function
\begin{equation}
\mathcal{L} \propto \prod_{i} \exp\bigg(\frac{(y_{GP}(x_i)-y_{mod}(x_i))^{2}}{2\sigma_{GP}^{2}(x_i)}\bigg),
\end{equation}
where $y_{GP}(x_i)$ are the values of the GP reconstructed function at the estimated data points, $y_{mod}(x_i)$ are the corresponding values of the function implied by the tested model and $\sigma_{GP}(x_i)$ are the uncertainties in the estimated data points.

One might argue that the reconstructed function may depend on the assumed kernel for the GP. However it was shown in \citet{MeliaYennapureddy2018,Mehrabi2022}, that using different kernels such as the above squared exponential kernel or the Matern kernel (v=3.5,4.5 and 92), does not lead to significant deviations. The main reason behind this is that the hyper-parameters used in the kernels are optimized for a given data set in such a way that the reconstructed function for each kernel is very similar. In this work we have used the squared exponential kernel for reconstructing the GP function for the Cosmic chronometers and HII Galaxy data sets.}
\subsection{Cosmic Chronometers}
Cosmic chronometers are luminous, red galaxies evolving on a timescale much larger than their age difference. These passively evolving galaxies are typically very massive with stellar material greater than $10^{11}$ M$_{\odot}$. Past observations have indicated that most massive galaxies seem to have an old stellar population by redshift $z \sim 1-2$~\citep{Dunlop1996,Spinrad1997,Cowie1999,Heavens2004,Thomas2005}, and that  90\% of their stellar mass has formed by redshift $z > 2$, with less than 1\% of stellar mass forming after redshift $z < 1$ \citep{Heavens2004,Panter2007}. Therefore these passively evolving galaxies are characterized by an old stellar population and a low star-formation rate (SFR). These passively evolving galaxies have a $4000 \mathrm{\AA}$ break in their spectra \citep{Moresco2011}, due to the metal absorption lines, with an amplitude linearly proportional to the metallicity and age difference of adjacent galaxies. Therefore given the metallicity one can infer the age difference ($\Delta t)$ from the amplitude difference of this $4000 \mathrm{\AA}$ break. Then using this age difference ($\Delta t$) and the inferred redshift difference, one can estimated $H(z)$ according to
\begin{equation}
    H(z) = \frac{-1}{1+z} \frac{dz}{dt} \approx \frac{-1}{1+z} \frac{\Delta z}{\Delta t}.
\end{equation}
This formulation is not based on any particular cosmology and the use of Gaussian Processes avoids the need for
fitting the data with predetermined parametric functions. Therefore this procedure provides a model-independent
measurement of $H(z)$.

This approach has been used by various workers to estimate $dz/dt$ and get $H(z)$ \citep{Stern2010a,Stern2010b, Moresco2012a,Moresco2012b}, whose model-independent values have then been used to test and constrain various cosmological models \citep{Yu2018,Busti2014,Li2016,Wang2017,MeliaYennapureddy2018}. Nevertheless, even this approach is subject to systematic uncertainties $\sigma_{\mathrm{sys}}$(see e.g. \citet{Moresco2012b}), arising from (1) a possible non-correlation arising from the change in age and stellar metallicity, or the star-formation history (SFH); (2) biasing due to the choice of a stellar population synthesis model that could affect the $H(z)$ estimate; and (3) the possibility that high-redshift early type galaxies might not be statistically equivalent to their low-redshift counterparts (i.e., a progenitor bias) \citep{vanDokkum1996}. In our case, we use the 30 measurements of $H(z)$ from \citet{leaf17} listed in Table 1. These were chosen from the compilation in \citet{zheng16} by omitting those values of $H(z)$ obtained with BAO. {The total errors for the calculated values of $H(z)$ in Table 1 include the statistical and systematic contributions $\sigma_{\mathrm{stat}}$ and $\sigma_{\mathrm{sys}}$ respectively. The main source for the systematic errors is derived from the stellar metallicity (\citep{MeliaYennapureddy2018}) and this is comparable to the statistical errors in the sample, while the progenitor bias and variations in the assumed star forming rates only contribute a few percent to $\sigma_{\mathrm{sys}}$. As has been done in previous studies that utilize the same sample of CC data (see for example \citet{nunes17} and \citet{leaf17} among others), the systematic errors in our case are assumed to be uncorrelated. This assumption may not be valid though, considering that metallicity variations with redshift may not be truly random (see \citet{lopez17}), and so it is safe to state that the systematic errors have correlated and uncorrelated components. In this case the total error may be written as \citep{MeliaYennapureddy2018}
\begin{equation}
\sigma(z_i) = \sqrt{\sigma_{\mathrm{stat}}^2 + f_s\sigma_{\mathrm{sys}}^2},
\end{equation}
where $f_s$ is a measure of the degree of correlation of the systematic errors, such that $f_s=0$ when these are fully correlated and $f_s = 1$ when they are totally random. As stated in \citet{MeliaYennapureddy2018} we do not know the value of $f_s$ or whether this is a constant or dependent on the redshift. Moreover only a subset of the 30 CC data points listed in Table 1 have separate values for $\sigma_{\mathrm{stat}}$ and $\sigma_{\mathrm{sys}}$ published. Hence the assumption that the systematic uncertainties are truly random means that the total errors in the sample shown in Table 1 may be over-estimated.} In the next section we shall reconstruct $H(z)$ by utilizing GP on these cosmic chronometers.
\begin{table}
 \centering
 \renewcommand{\arraystretch}{0.7}
 \begin{tabular}{ccr}
 \hline
 $z$ & $H(z)\ \mathrm{(km s^{-1}Mpc^{-1})}$ & References \\
 \hline
 $0.09$ & $69\pm12$ & \citep{jimenez03}\\
 \hline
 $0.17$ & $83\pm8$ & \citep{simon05}\\
 $0.27$ & $77\pm14$ & \\
 $0.4$ & $95\pm17$ & \\
 $0.9$ & $117\pm23$ & \\
 $1.3$ & $168\pm17$ & \\
 $1.43$ & $177\pm18$ & \\
 $1.53$ & $140\pm14$ & \\
 $1.75$ & $202\pm40$ & \\
 \hline
 $0.48$ & $97\pm62$ &\citep{Stern2010b}\\
 $0.88$ & $90\pm40$ &\\
 \hline
 $0.1791$ & $75\pm4$ & \citep{Moresco2012b}\\
 $0.1993$ & $75\pm5$ & \\
 $0.3519$ & $83\pm14$ & \\
 $0.5929$ & $104\pm13$ & \\
 $0.6797$ & $92\pm8$ & \\
 $0.7812$ & $105\pm12$ & \\
 $0.8754$ & $125\pm17$ & \\
 $1.037$ & $154\pm20$ & \\
 \hline
 $0.07$ & $69\pm19.6$ & \citep{zhang14}\\
 $0.12$ & $68.6\pm26.2$ & \\
 $0.2$ & $72.9\pm29.6$ & \\
 $0.28$ & $88.8\pm36.6$ & \\
 \hline
 $1.363$ & $160\pm33.6$ & \citep{moresco15} \\
 $1.965$ & $186.5\pm50.4$ & \\
 \hline
 $0.3802$ & $83\pm13.5$ & \citep{moresco16}\\
 $0.4004$ & $77\pm10.2$ & \\
 $0.4247$ & $87.1\pm11.2$ & \\
 $0.4497$ & $92.8\pm12.9$ & \\
 $0.4783$ & $80.9\pm9$ & \\
 \hline
 \end{tabular}
 \caption{Hubble parameter $H(z)$ from cosmic chronometers.}
\end{table}

\subsection{HII Galaxies}
HII galaxies (HIIGx) are massive starburst structures surrounded by ionized hydrogen gas. Their luminosity is mostly dominated by the starbursts. Due to the presence of ionized hydrogen gas surrounding them, the optical spectra of these galaxies include strong Balmer emission lines in H$\alpha$ and H$\beta$. Giant extragalactic HII regions (GEHR) located mostly in the outskirts of galaxies share very similar characteristics. In particular these regions also have a high star formation rate and massive star clusters surrounded by ionized hydrogen gas. Hence both the HIIGx and GEHR have very similar optical spectra with strong Balmer emission lines in H$\alpha$ and H$\beta$ \citep{Searle1972, Bergeron1977, Terlevich1981, Kunth2000}. The relevance of these sources to the work described here is that their luminosity L(H$\beta$) in H$\beta$ and the ionized gas velocity dispersion $\sigma(H\beta)$ \citep{Terlevich1981} in these systems are strongly correlated. This happens because the intensity of ionizing radiation and velocity gas dispersion both increase with the starburst mass. Moreover, the relatively small scatter in the L(H$\beta$) versus $\sigma(H\beta)$ relation makes these sources ideal standard candles \citep{Melnick1987, Melnick1988, Fuente2000, Melnick2000, Bosch2002, Telles2003, Bordalo2011, Plionis2011, Mania2012, Chavez2012,Chavez2014, Terlevich2015}. By using this relationship, one can infer the luminosity distance  which can in turn be used to obtain the Hubble constant $H_0$ \citep{Melnick1988} and also constrain various cosmological models \citep{Wei2016,Yennapureddy2017}.

{In this paper, we use data from \citet{Gonzalez2021} consisting of the VLT-KMOS high spectral resolution observations of 41 high-z $(1.3 \leq z \leq 2.6)$ HIIGx, together with other published data \citep{Chavez2014,Terlevich2015,Fernandez2018,Gonzalez2019} for 45 high-z and 107 $(z \leq 0.15)$ HIIGx.} The luminosity L(H$\beta$) versus $\sigma(H\beta)$ relationship is given as \citep{Chavez2012}
\begin{equation}
    \log \left[\frac{L(H\beta)}{\mathrm{erg\,s^{-1}}}\right] = \alpha \log \left[\frac{\sigma (H\beta)}{\mathrm{km\,s^{-1}}}\right] + \kappa, \label{Lsigma}
\end{equation}
where $\alpha$ and $\kappa$ are constants representing the slope and intercept.
{The selection of HII galaxy sample has been done according to a set of criteria (see \citet{Terlevich2015} and \citet{Ruan19} for details) such that the bolometric flux of the HIIGx is regarded to comprise mainly the H$\beta$ line. In this case the luminosity distance of an HIIGx is given approximately by \begin{equation}
d_L = \left[\frac{L(\mathrm{H}\beta)}{4\pi F(\mathrm{H}\beta)}\right]^{1/2},
\end{equation}
where $L(\mathrm{H}\beta)$ and $F(\mathrm{H}\beta)$ are the luminosity and flux associated with the H$\beta$ line. Taking the log of both sides of the above equations and using (\ref{Lsigma}) leads to
\begin{equation}
\log\left(\frac{d_L}{\mathrm{Mpc}}\right) = 0.5\left[\alpha\log\left(\frac{\sigma(\mathrm{H}\beta)}{\mathrm{km\, s^{-1}}}\right) - \log\left(\frac{F(\mathrm{H}\beta)}{\mathrm{erg\,s^{-1}\,cm^{-2}}}\right)\right] + 0.5\kappa  - 25.04. \label{lumHb}
\end{equation}
Then defining the parameter $\delta$ by
\begin{equation}
\delta \equiv -2.5\kappa - 5\log \left[\frac{H_0}{\mathrm{km\,s^{-1}\,Mpc^{-1}}}\right] + 125.2,
\end{equation}
one can write (\ref{lumHb}) in terms of the dimensionless luminosity distance $(H_0 d_L)/c$ as
\begin{equation}
\left(\frac{H_0}{c}\right) d_{L}(z) = \frac{10^{\eta(z)/5}}{c/(\mathrm{km\,s^{-1}})},
\end{equation}
where
\begin{equation}
\eta(z) = -\delta + 2.5\left(\alpha\log\left[\frac{\sigma(\mathrm{H}\beta)}{\mathrm{km\,s^{-1}}}\right] - \log\left[\frac{F(\mathrm{H}\beta)}{\mathrm{erg\,s^{-1}\,cm^{-2}}}\right]\right). \label{eta}
\end{equation}
The function $\eta(z)$ is related to the distance modulus $\mu(z)$ by
\begin{equation}
\mu(z) = \eta(z) - 5\log\left(\frac{H_0}{\mathrm{km\,s^{-1}\,Mpc^{-1}}}\right) + 25. \label{obs-dm}
\end{equation}
Therefore from the flux $F(\mathrm{H}\beta)$ and gas velocity dispersion $\sigma(\mathrm{H}\beta)$, one can use (\ref{eta}) and (\ref{obs-dm}) to calculate the distance modulus for the HIIGx and GEHR.
In principle, one has to optimize the parameters $\alpha$ and $\delta$ simultaneously with the cosmological parameters to avoid circularity issues, but past studies \citep{Wei2016} have shown that $\alpha$ and $\delta$ are negligibly sensitive to the cosmological model and their values change by at most a tiny fraction of their $1\sigma$ errors irrespective of the model being tested. So instead of optimizing $\alpha$ and $\delta$ for our models we use the average optimized values for these parameters used in \citet{Wei2016}. These are $\alpha=4.87^{+0.11}_{-0.08}$, $\delta=32.42^{+0.42}_{-0.33}$.} Correspondingly the theoretical distance modulus is given by
\begin{equation}
    \mu_{th}=5\log\bigg[\frac{d_L(z)}{\mathrm{Mpc}}\bigg] + 25, \label{distmodlum}
\end{equation}
where $d_L(z)$ is the luminosity distance for each f(R) model given by (\ref{lumdist}). In our case, the luminosity distance is extracted via a GP reconstruction of the HIIGx Hubble diagram and this is then used together with $H(z)$ to perform an independent, as well as a joint, analysis of the $f(R)$ models as described in the next section.\\
{Finally as in the case of the CC the HII galaxy probe based on the $L(\mathrm{H}\beta)-\sigma$ relation is also subject to systematic uncertainties. These are attributed to the size and age of the starburst, the oxygen abundance of the HIIGx and the internal extinction correction \citep{Chavez2016}. Although most of the scatter in the $L-\sigma$ relation is due to observational errors \citep{Melnick2000} these uncertainties still need to be better understood.}
\section{Results}
We shall begin by obtaining constraints on the $f(R)$ models based solely on the cosmic chronometer observations. After reconstructing the expansion rate $H(z)$ using Gaussian Processes on these data (Table 1), as described in the previous section, we optimize the fit for each of the three $f(R)$ models using the analytic approximations for $H(z)$ given in Equations~(\ref{approx1}-\ref{approx3}). This is shown in Figure 1, in which the GP reconstructed $H(z)$ is shown as a solid black curve and the fitted approximated function for each model is shown as a dotted blue curve. The shaded part represents the $1\sigma$ confidence region. We then use the HII galaxy data and obtain a GP reconstruction of the distance modulus $\mu(z)$ as given by Equation~(\ref{obs-dm}), using the values of the parameters $\alpha$ and $\delta$ mentioned in the previous section. Since the distance modulus is related to the luminosity distance via Equation~(\ref{distmodlum}), we then use the approximate analytic expression for the luminosity distance given in Equation~(\ref{lumdist}) to obtain an optimized fit for each $f(R)$ model. This is shown in Figure 2, where again the continuous black curve is the GP reconstructed $\mu(z)$ and the dotted blue curve is the fitted function, with the shaded part representing the $1\sigma$ confidence region. This is followed by a joint analysis using both the cosmic chronometers and HII galaxy data together to obtain constraints for the deviation parameter $b$ and the present values of the matter density parameter $\Omega_m$ and dimensionless Hubble parameter $h$ for each model. The best fit values for these parameters from the cosmic-chronometer data and the joint analysis are shown in Table 2. The superimposed contour plots for the cosmic-chronometer and joint analysis data sets in all three models are shown in Figure 3.

{From Table 2 it can be seen clearly that the optimized values for the model parameters and their errors in
the CC + HIIGx joint analysis (second column) are very similar to those obtained when the CC data are used
alone (first column).
So in order to get a better understanding of this result, we have obtained the optimized values of
the model parameters using the HIIGx data alone, and in this case we found that the error bars for these
parameters are very large compared to those obtained with the CC data alone. This can be explained by the fact
that $\mu(z)$, being an integrated quantity unlike $H(z)$, is less responsive to the values of the model
parameters, and therefore provides weaker constraints on the values of the optimized parameters when used on
their own. So taking the ArcTanh model as an example, in Figure 4 we have plotted $\mu(z)$ for different values
of $b$ and $\Omega_m$ using Equations (\ref{distmodlum}) and (\ref{lumdist}). In this case the changes in $\mu(z)$ are minimal, particularly when the deviation parameter $b$ is changed.} {This would
also have happened even if we had used other cosmological data such as SNeIa observations, and so it
wouldn't make sense to extend the analysis by including SNeIa data.}

\captionsetup[subfigure]{labelformat=empty}
\begin{figure}
\centering
\subfloat[(a)]{\includegraphics[width=.45\linewidth]{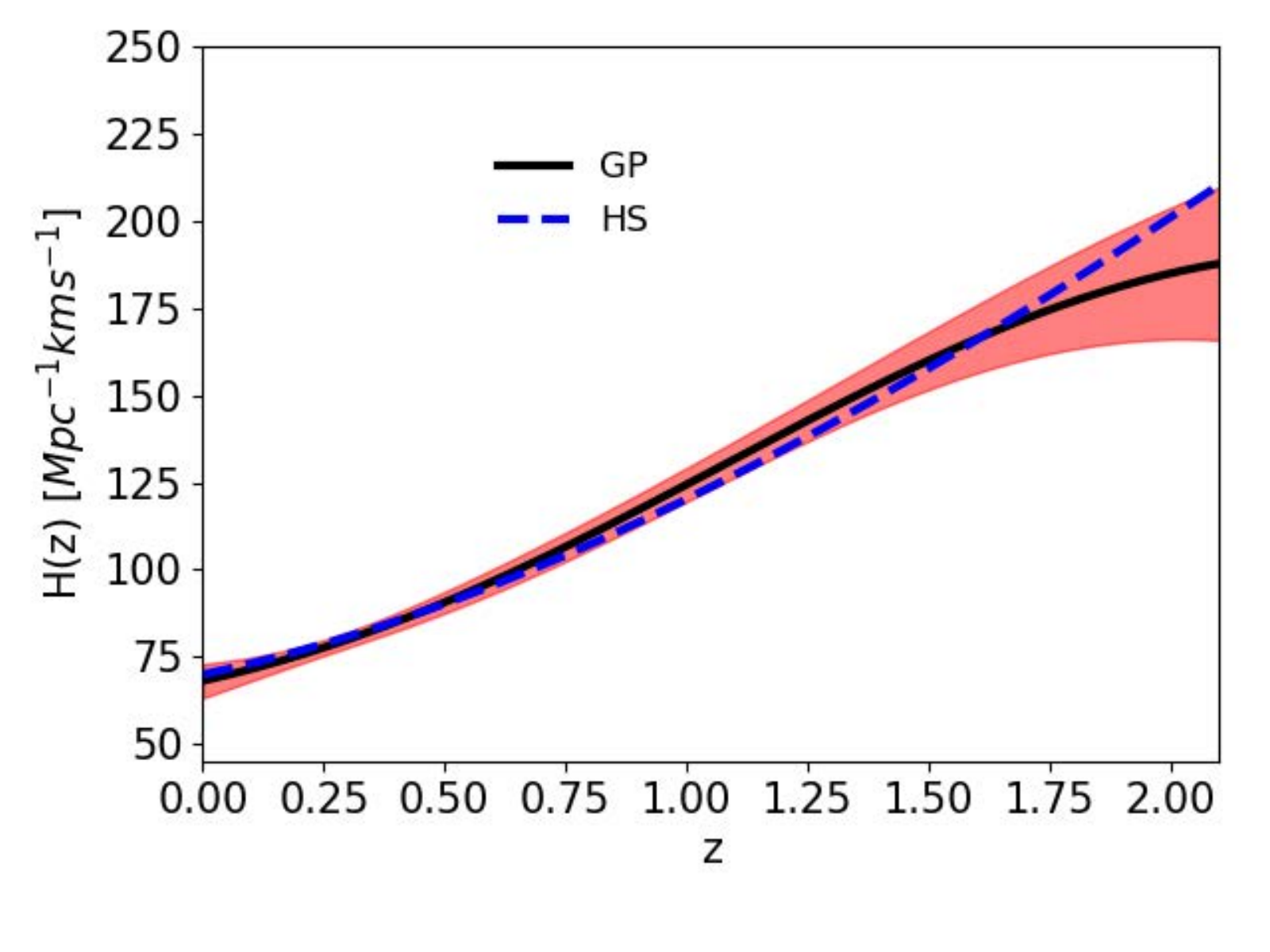}}
\hskip 0.2in
\subfloat[(b)]{\includegraphics[width=.45\linewidth]{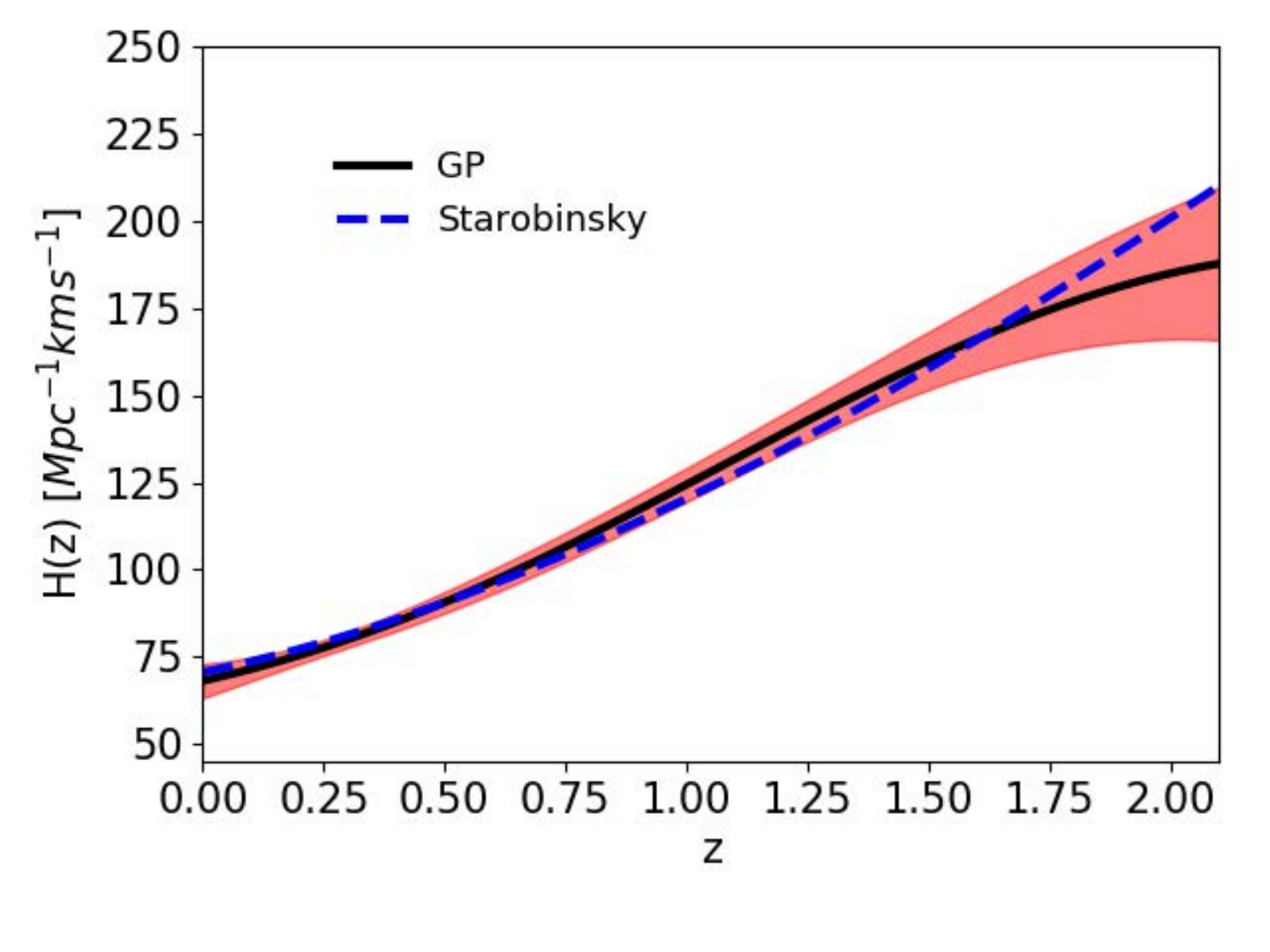}}
\vspace{0.5cm}
\subfloat[(c)]{\includegraphics[width=.45\linewidth]{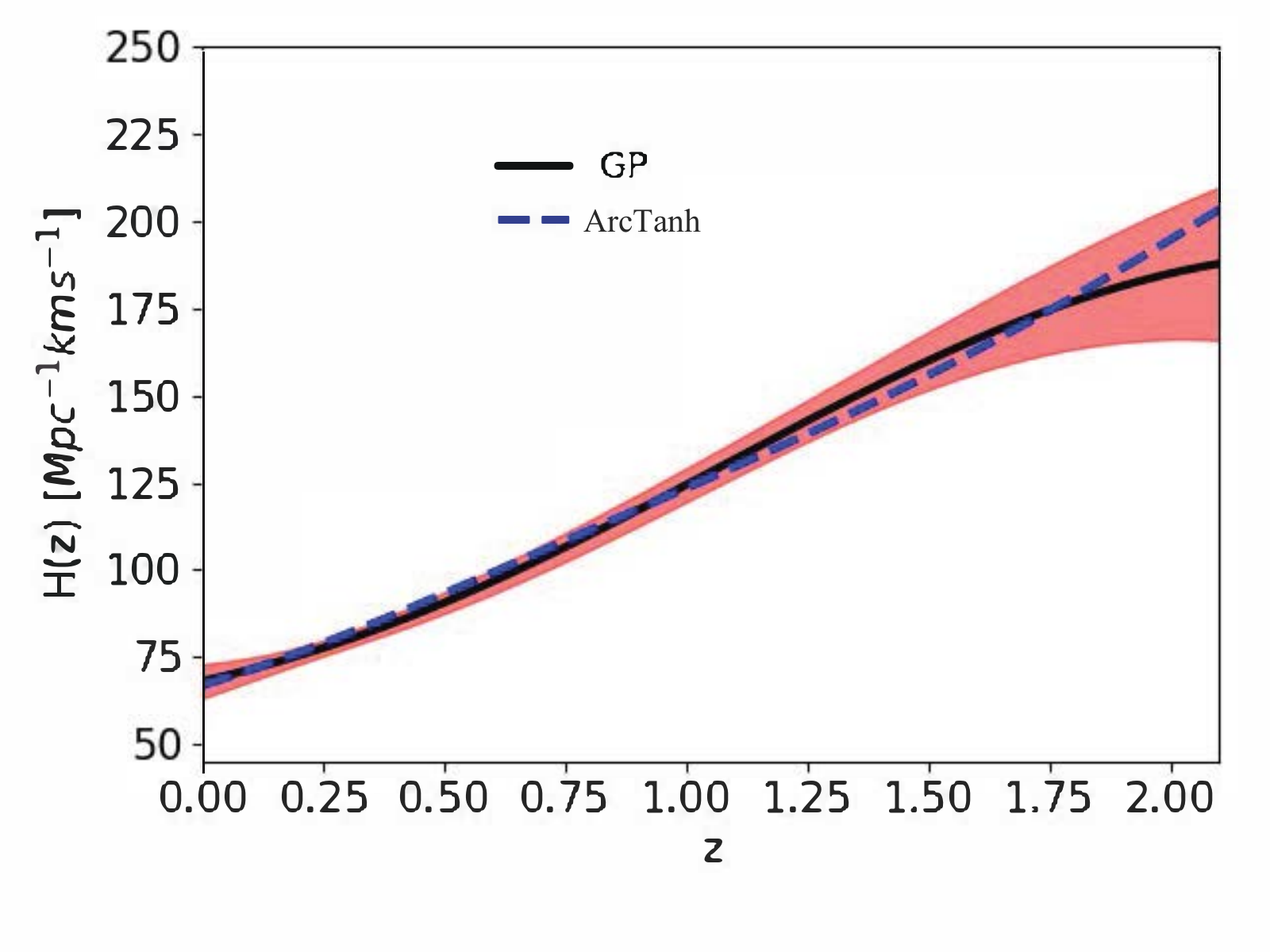}}
\caption{The Hubble parameter $H(z)$ (blue dashed curve) as given in Equations~(\ref{approx1}-\ref{approx3}) optimized to fit the GP reconstructed function (black solid curve) from the cosmic-chronometer data for (a) the HS-model, (b) the Starobinsky model and (c) the Arctanh model. The red band represents the $1\sigma$ confidence region.}
\end{figure}

\captionsetup[subfigure]{labelformat=empty}
\begin{figure}
\centering
\subfloat[(a)]{\includegraphics[width=.45\linewidth]{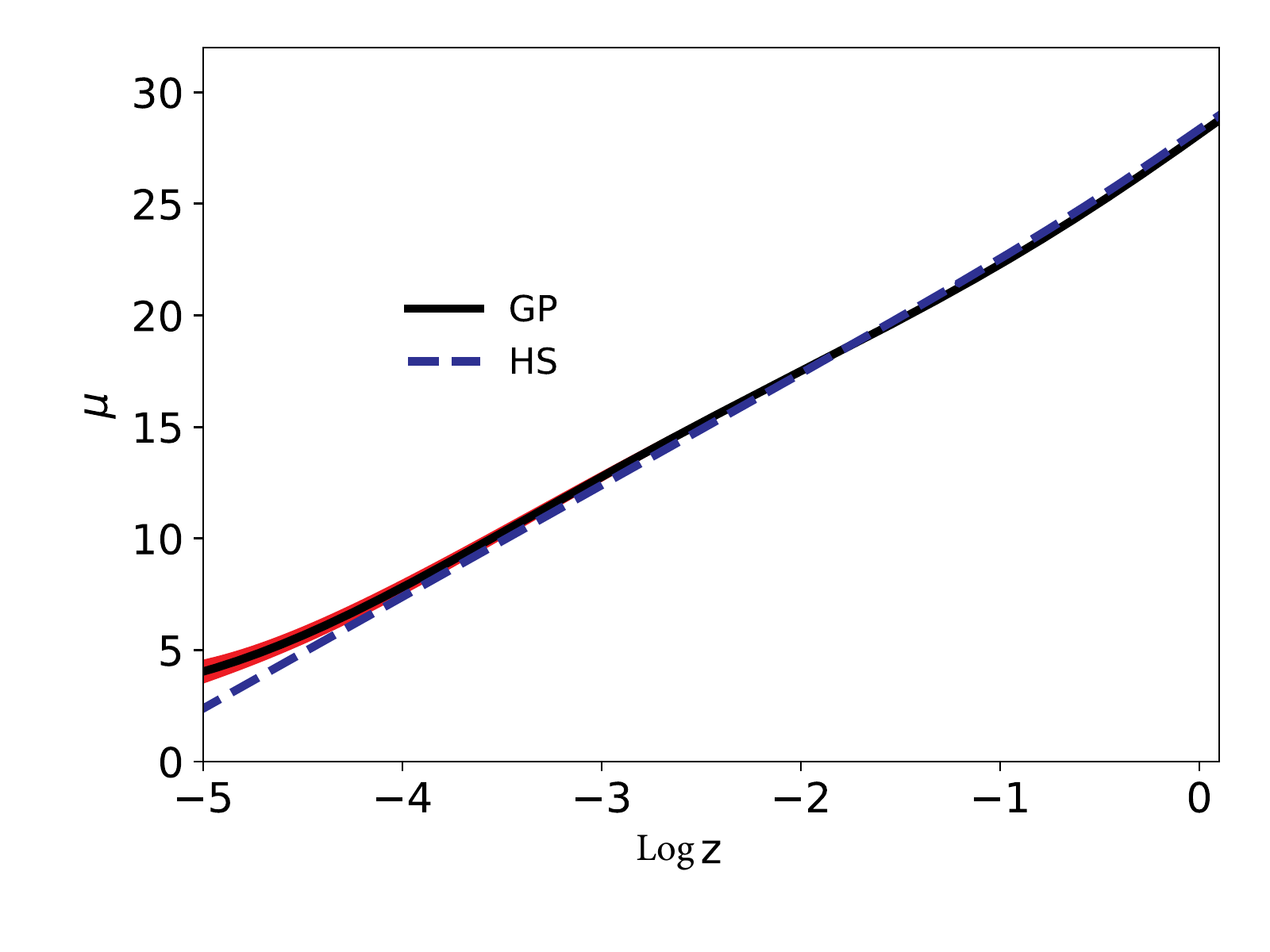}}
\hskip 0.2in
\subfloat[(b)]{\includegraphics[width=.45\linewidth]{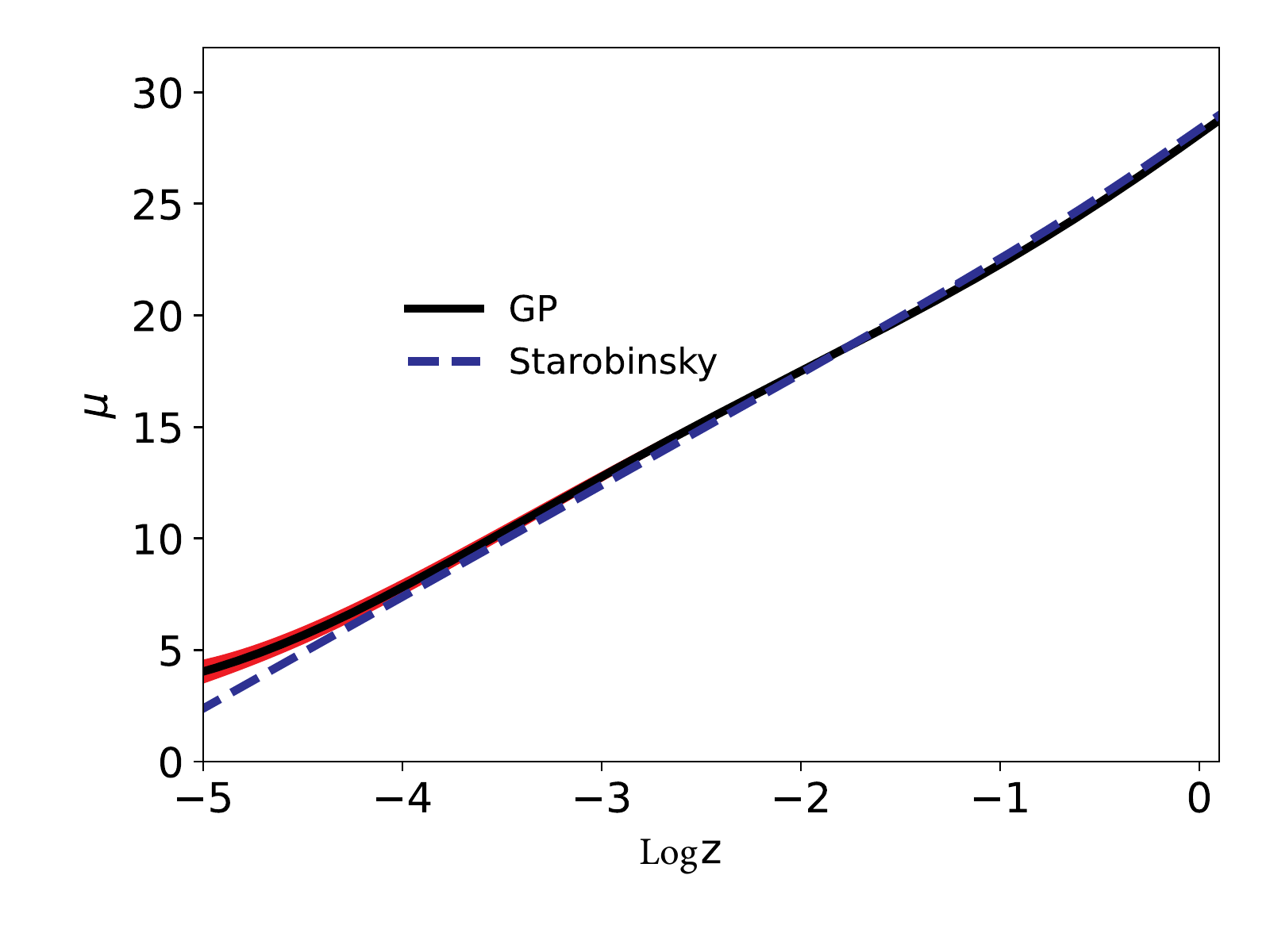}}
\vspace{0.5cm}
\subfloat[(c)]{\includegraphics[width=.45\linewidth]{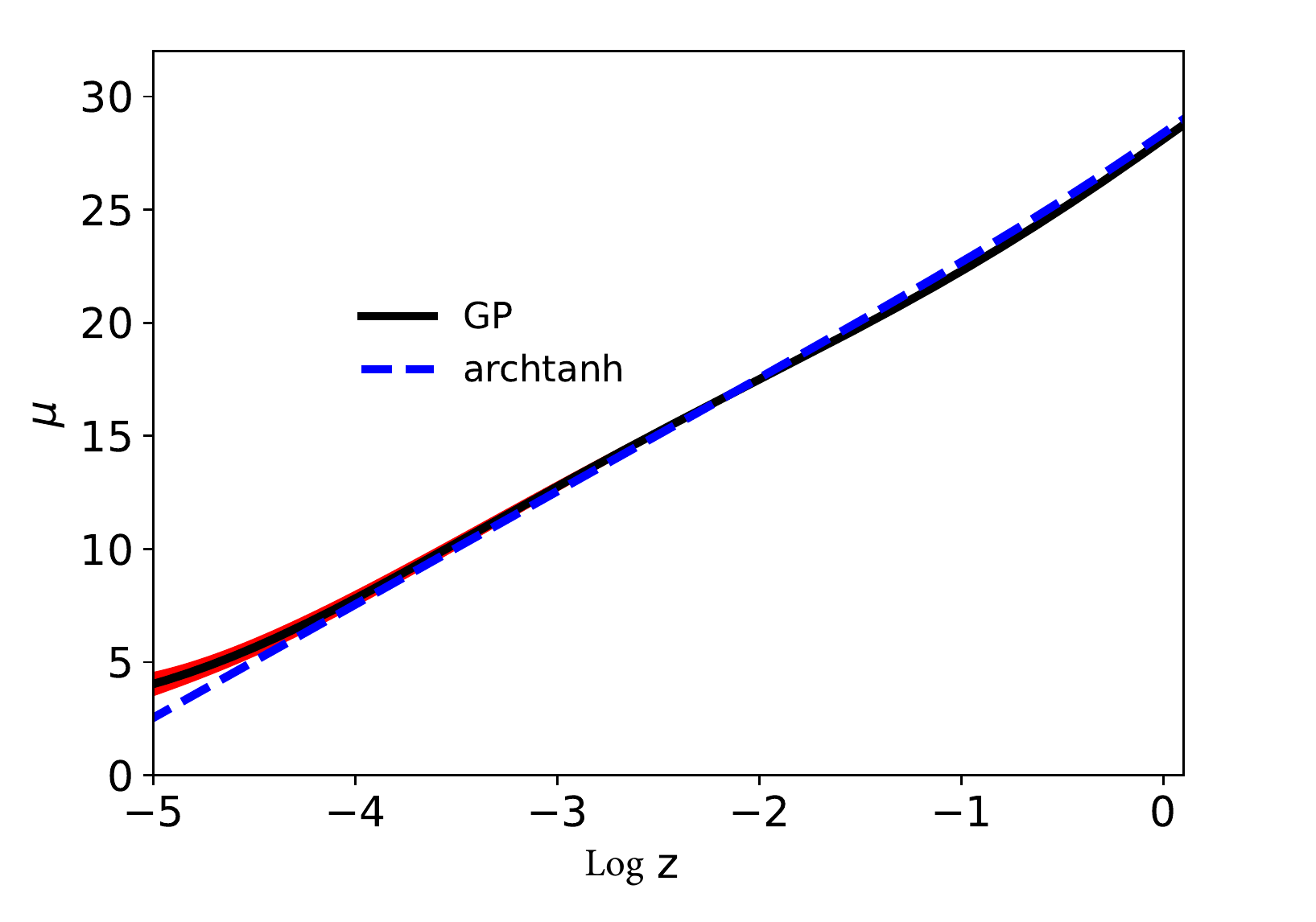}}
\caption{The distance modulus $\mu(z)$ (blue dashed curve) optimized to fit the GP reconstructed function (black solid curve) for the HII galaxy data, for (a) the HS-model, (b) the Starobinsky model and (c) the Arctanh model. The red band represents the $1\sigma$ confidence region.}
\end{figure}

\captionsetup[subfigure]{labelformat=empty}
\begin{figure}
\centering
\subfloat[(a)]{\includegraphics[width=.45\linewidth]{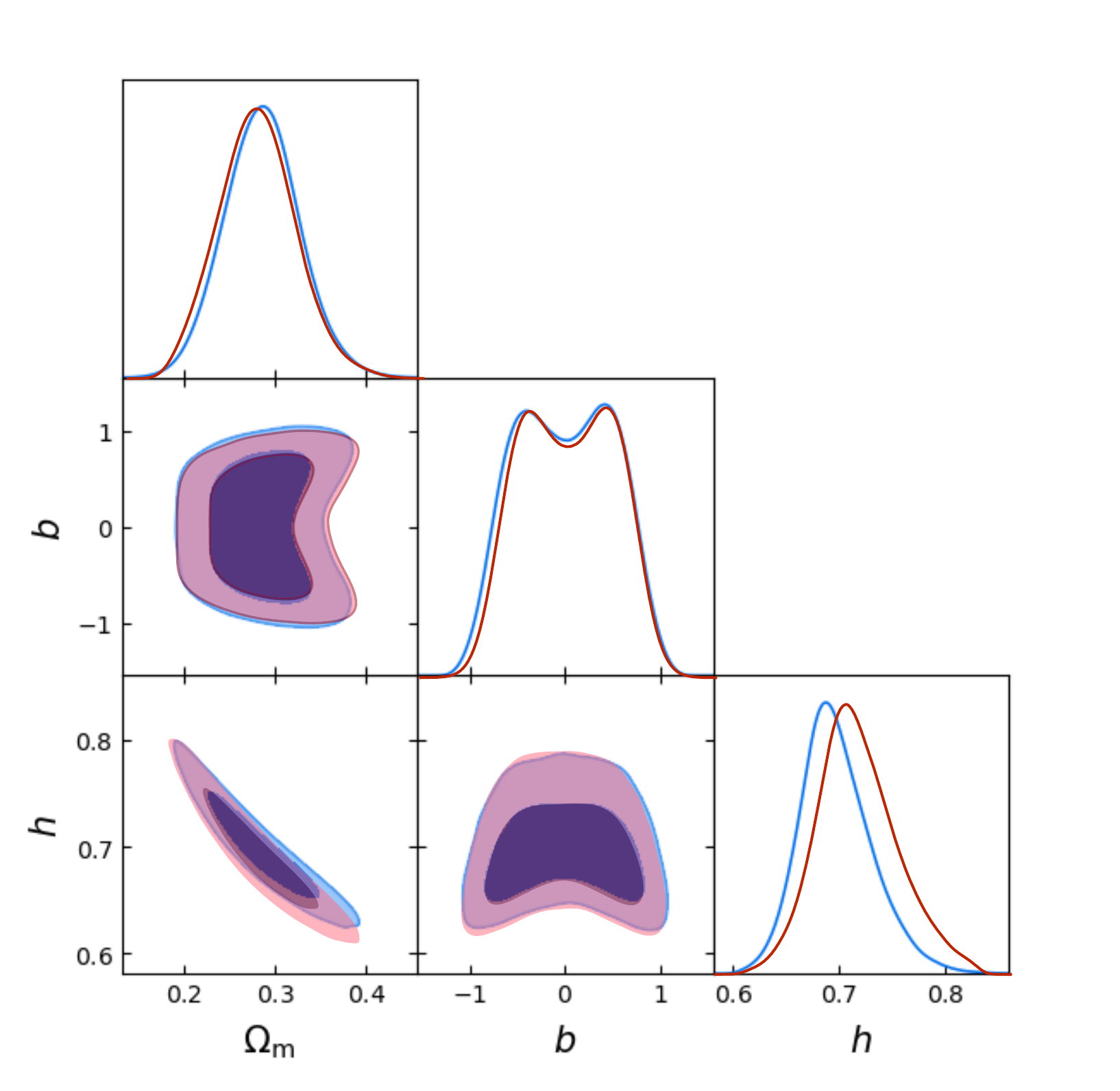}}
\hskip 0.2in
\subfloat[(b)]{\includegraphics[width=.45\linewidth]{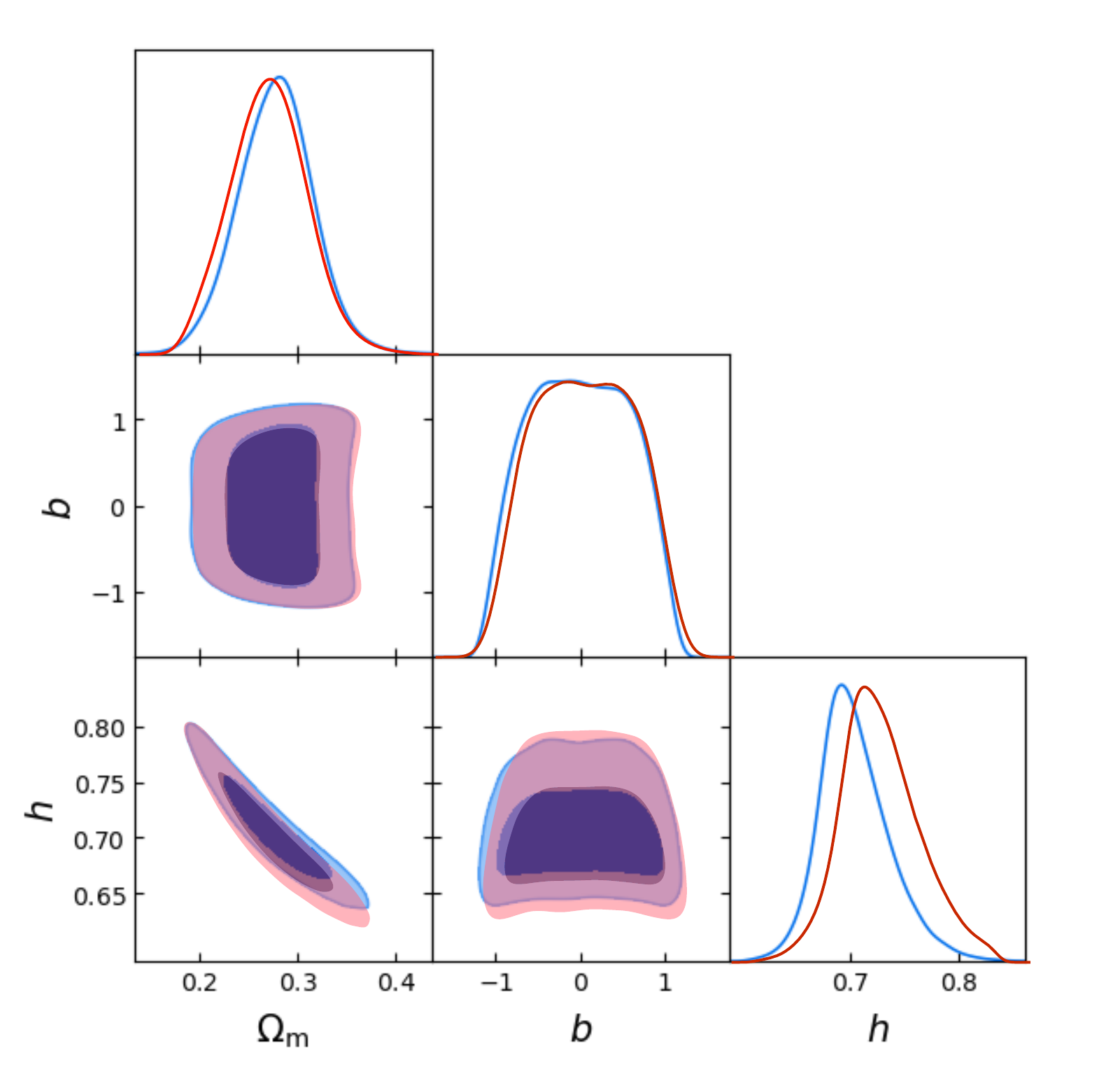}}
\vspace{0.5cm}
\subfloat[(c)]{\includegraphics[width=.45\linewidth]{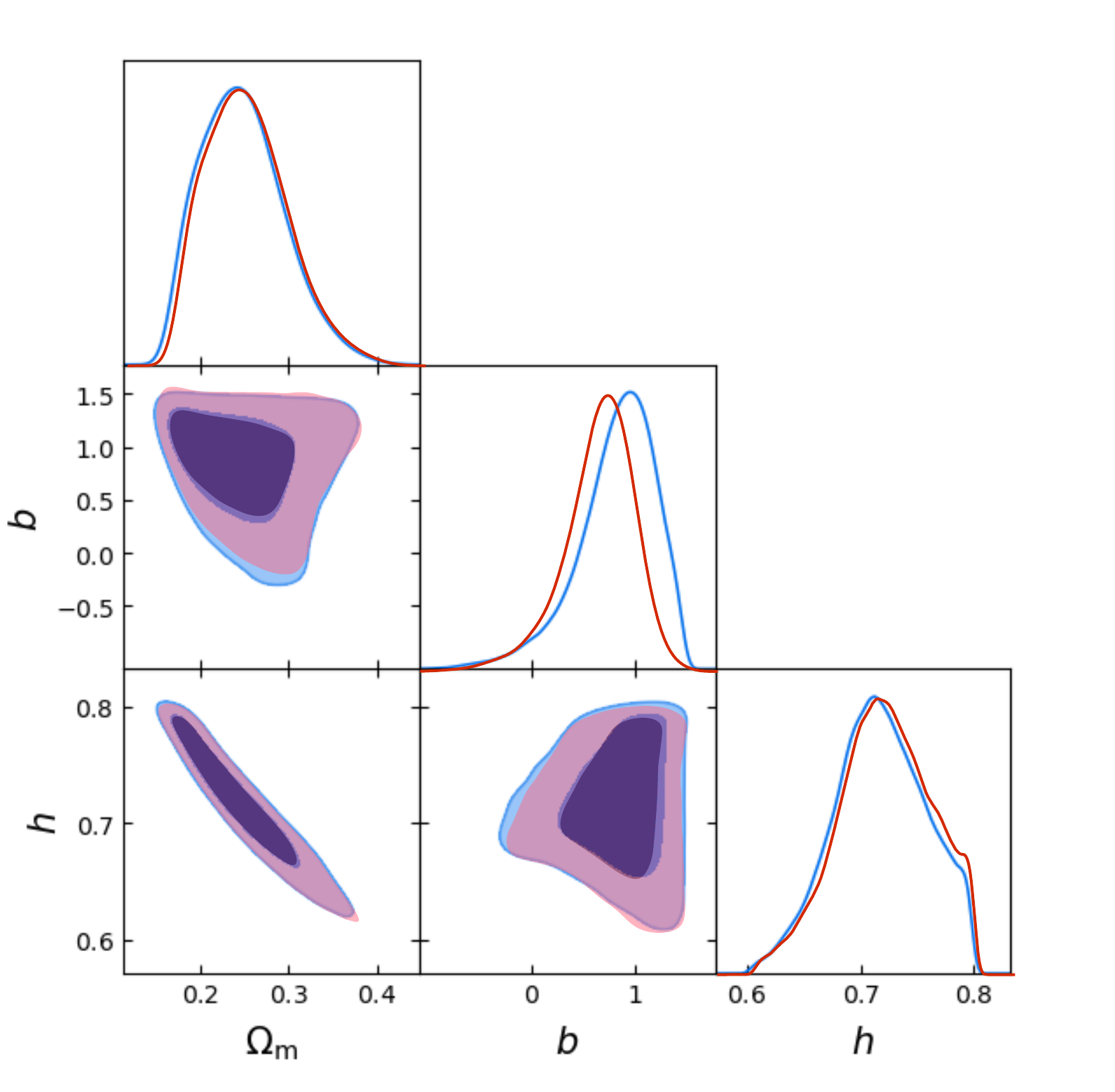}}
\caption{Contour plots for the model parameters $b$, $\Omega_m$  and $h$  for (a) the HS-model, (b) the Starobinsky model and (c) the Arctanh model. Red corresponds to the cosmic chronometers (CC) data set, while blue corresponds to the CC + HIIGx joint analysis. The darker and lighter shades represent the $1\sigma$ and $2\sigma$ confidence regions.}
\end{figure}

\captionsetup[subfigure]{labelformat=empty}
\begin{figure}
\centering
\subfloat[(a)]{\includegraphics[width=.45\linewidth]{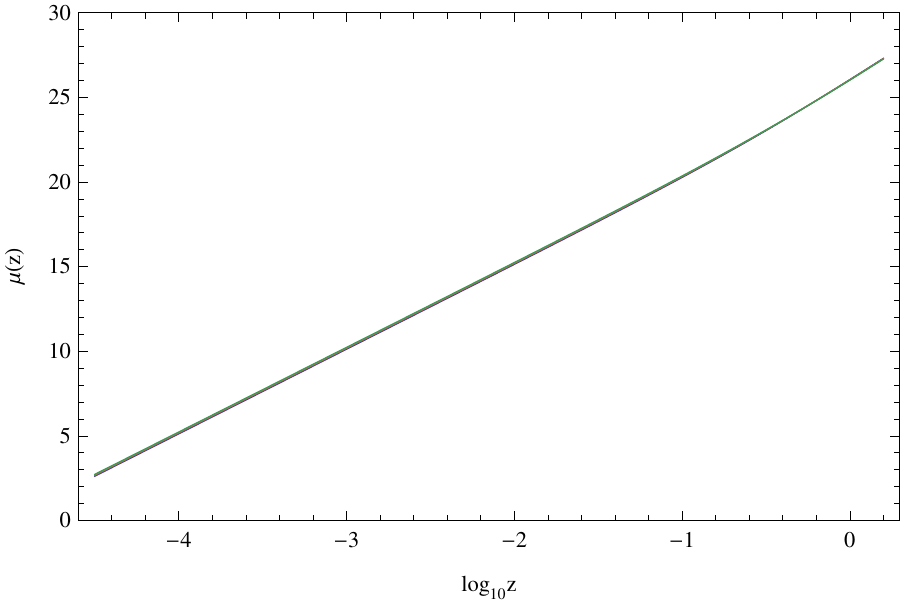}}
\hskip 0.2in
\subfloat[(b)]{\includegraphics[width=.45\linewidth]{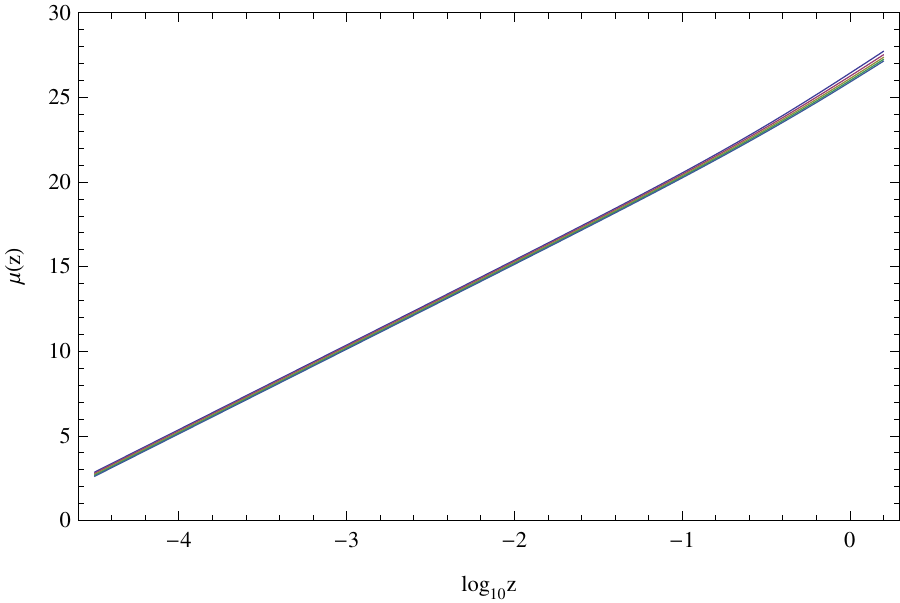}}
\caption{The distance modulus $\mu(z)$ obtained from Equations (\ref{distmodlum}) and (\ref{lumdist}) for (a) different values of $b$ in the range $b \in [0.5, 1.2]$ with $\Omega_m = 0.231$, (b) different values of $\Omega_m$ in the range $\Omega_m \in [0.1,0.3]$ (top curve corresponds to $\Omega_m = 0.1$) with $b=0.901$.}
\end{figure}

\begin{table}
 \centering
 \begin{tabular}{|l|c|c|c|}
 \hline
 Model & \;\;Parameters\;\; & CC & CC + HII \\
 \hline
 & $b$ & \;\;$0.003^{+0.528}_{-0.533}$\;\; & \;\;$0.003^{+0.559}_{-0.563}$\;\;\\[2mm]
 {HS}&$\Omega_{m}$ & $0.279^{+0.040}_{-0.039}$ & $0.286^{+0.041}_{-0.041}$\\[2mm]
 & $h$ & $0.699^{+0.037}_{-0.029}$ & $0.694^{+0.037}_{-0.029}$ \\
 \hline
 & $b$ & $0.006^{+0.632}_{-0.635}$ & $-0.004^{+0.650}_{-0.641}$\\[2mm]
 {Starobinsky}&$\Omega_{m}$ & $0.273^{+0.037}_{-0.037}$ & $0.278^{+0.036}_{-0.038}$\\[2mm]
 & $h$ & $0.706^{+0.036}_{-0.027}$ & $0.702^{+0.035}_{-0.026}$ \\
 \hline
 & $b$ & $0.901^{+0.318}_{-0.397}$ & $0.872^{+0.315}_{-0.408}$\\[2mm]
 {ArcTanh}&$\Omega_{m}$ & $0.231^{+0.051}_{-0.043}$ & $0.245^{+0.051}_{-0.046}$\\[2mm]
 & $h$ & $0.729^{+0.044}_{-0.042}$ & $0.717^{+0.044}_{-0.040}$ \\
 \hline
 \end{tabular}
\caption{Best fit values for the model parameters $b$, $\Omega_m$  and $h$ using cosmic chronometers (CC) and CC + HII galaxy joint analysis.}
\end{table}

\section{Model Comparisons}
After optimizing the model parameters by fitting the data, we next examine which of them are favoured by the observations. To do so we use standard information criteria, namely the Akaike Information Criterion (AIC) \citep{aic} and the Bayesian Information Criterion (BIC) \citep{bic}. Besides the value of $\chi^2_{\mbox{min}}$, these criteria also take into consideration the number of free parameters, and so are useful when comparing models that are not nested. The AIC is defined by
\begin{equation}
\mathrm{AIC} = \chi^2_{\mbox{min}} + 2d,
\end{equation}
where $d$ is the number of free parameters in the model. The BIC is similarly defined by
\begin{equation}
\mathrm{BIC} = \chi^2_{\mbox{min}} + d\ln N,
\end{equation}
where $N$ is the number of data points used in the analysis. In general, a smaller AIC/BIC indicates a greater consistency with the data. In order to access which of the different models is favoured by the given measurements, one normally calculates the differences $\Delta\mathrm{AIC} = \mathrm{AIC}_2 - \mathrm{AIC}_1$ and $\Delta \mathrm{BIC} = \mathrm{BIC}_2 - \mathrm{BIC}_1$ such that the higher the value of $|\Delta \mathrm{AIC}|$ ($|\Delta \mathrm{BIC}|$), the greater the evidence against the model with the higher value of AIC (BIC). The result is considered ``positive'' when $2 \leq |\Delta \mathrm{AIC}| (|\Delta \mathrm{BIC}|) \leq 6$, ``strong'' when $6 \leq |\Delta \mathrm{AIC}| (|\Delta \mathrm{BIC}|) \leq 10$ and ``very strong'' when $|\Delta \mathrm{AIC}| (|\Delta \mathrm{BIC}|) \geq 10$. A difference of less than 2 indicates that the two models being compared are statistically indistinguishable.
\begin{table}
 \centering
 \begin{tabular}{|l|c|c|c|c|c|c|c|c|}
 \hline
 \multirow{2}{3cm}{Model} & \multicolumn{4}{c|}{CC} & \multicolumn{4}{c|}{CC+HII} \\
 \cline{2-9}
 & AIC & $\Delta$AIC & BIC & $\Delta$BIC & AIC & $\Delta$AIC & BIC & $\Delta$BIC\\
 \hline
 $\Lambda$CDM & 17.64 & 0 & 27.53 & 0 & 27.46 & 0 & 37.35 & 0 \\
 HS & 23.04 & 5.40 & 32.93 & 5.40 & 24.16 & -3.3 & 34.05 & -3.3 \\
 Starobinsky & 22.96 & 5.32 & 32.85 & 5.32 & 24.10 & -3.36 & 33.99 & -3.36 \\
 ArcTanh & 18.36 & 0.72 & 28.25 & 0.72 & 20.34 & -7.12 & 30.23 & -7.12 \\
 \hline
 \end{tabular}
\caption{The AIC and BIC values for the CC and CC+HII galaxy data sets for the three analyzed $f(R)$ models and $\Lambda$CDM.}
\end{table}

The AIC and BIC values for our models and $\Lambda$CDM are shown in Table 3, for both the CC and combined CC+HII galaxy data sets. We also include the differences $\Delta$AIC and $\Delta$BIC between the AIC(BIC) values of the $f(R)$ models and $\Lambda$CDM. These will indicate how these models compare with standard $\Lambda$CDM.

It is clear from the above analysis that the HS and Starobinsky models are consistent with each other using either the CC data on their own, or in combination with the HII galaxy measurements. Based solely on the CC data, $\Lambda$CDM fares better than all three $f(R)$ models, particular the HS and Starobinsky models. The ArcTanh model and $\Lambda$CDM are statistically indistinguishable in this case. The situation changes completely when the combined CC and HII galaxy data are used, however. In this case, there is positive evidence that the ArcTanh model is favored over $\Lambda$CDM.

\section{Discussion and Conclusion}
In this paper, we have used cosmic chronometers and HII galaxies to constrain the parameters in several viable $f(R)$ models in the literature, including the two best know ones by Hu \& Sawicki and Starobinsky, and the novel ArcTanh model introduced recently in \citet{romero18}. These and other viable $f(R)$ models have been analyzed extensively in the past using different individual and combined data sets. Unlike previous studies, however, the method we have used to obtain these constraints does not rely on any fiducial cosmology or assumed priors. For example, recently Nunes et al. \citep{nunes17} have also used cosmic chronometers to constraint some $f(R)$ models, but they did not use GP to extract $H(z)$ from the data and the fitting for each model was done by numerically integrating the Friedman equation. Moreover, in their study they used a combined data set, including BAO measurements that. { As already mentioned, the use of BAO is highly model dependent, considering that in general in order to disentangle the redshift space distortions from the actual redshift measurements, one needs to use a fiducial model (taken to be a basic version of $\Lambda$CDM) to simulate these distortions. Therefore we decided not to include BAO data in our analysis.}

We have used GP to reconstruct $H(z)$ from cosmic chronometers and the luminosity distance $d_L(z)$ from the HII galaxy Hubble diagram. As is the case for cosmic chronometers, the parametrization of HII galaxies given in Equation~(\ref{obs-dm}) is known to be insensitive to the underlying cosmology. Moreover, the use of GP avoids the need to fit the data to a pre-determined parametrization indicated by a particular model. We have obtained analytic series approximations for $H(z)$ and $d_L(z)$ for the $f(R)$-models in terms of the free parameters $b$, $\Omega_m$ and $h$, which are then fitted to the corresponding GP-derived functions to obtain the optimized values of these parameters for each model.

The fitting was first carried out with $H(z)$ from cosmic chronometers independently and then this was followed by a joint analysis using both $H(z)$ and $d_L(z)$ from cosmic chronometers and HII galaxies. This approach tests whether the joint analysis would lead to different and tighter constraints for the model parameters. In fact, unlike previous studies \citep{nunes17,romero18} which in general exhibited significantly different and tighter constraints for the deviation parameter $b$ when combined data sets were used, our optimized values for $b$ and the corresponding $1\sigma$ confidence intervals for all three models did not change significantly when the joint analysis was performed. This applies also for the other optimized values of $\Omega_m$ and $h$, as can be seen in the contour plots for the three models in Figure 3, where the associated $1 \sigma$ and $2\sigma$ confidence regions are almost overlapping. {As mentioned in Section IV, this occurs due to the fact that when the HIIGx data are used on their own, the obtained constraints for the model parameters are not as good as those obtained from the CC data.}

For the HS and Starobinsky models, the optimized value of the deviation parameter is much closer to zero than what has been reported in the literature. In both cases, the zero value for $b$ falls safely within the $1\sigma$ confidence levels, which indicates that these models coincide observationally with the predictions of $\Lambda$CDM. This is not so for the ArcTanh model, however, where the optimized value of $b$ is clearly non-zero. The zero value of $b$ in this case does not even lie in the $1\sigma$ confidence interval, which indicates that this model deviates appreciably from $\Lambda$CDM. It should be noted that in the recent study of \citet{romero18}, where this particular model was introduced along with others, the optimized value of the deviation parameter for this model, from growth rate data alone, was inferred to be $b = 2\pm2$, decreasing to $b = 0.010\pm0.007$ when a combined data set, including Type Ia SNe, BAO and the CMB, was used. An analysis of the AIC and BIC reveals that there is no statistical distinction between the HS and Starobinsky models for either data sets. When the CC data are used on their own, $\Lambda$CDM does better than all three $f(R)$ models, particularly HS and Starobinsky. On the other hand, when the combined CC+HII galaxy data are used, there is positive evidence favouring ArcTanh over $\Lambda$CDM.

\section*{Data Availability}
The data used in this article are available in \citet{leaf17} and \citet{Chavez2014,Terlevich2015,Fernandez2018,Gonzalez2019,Gonzalez2021}.


\bsp

\label{lastpage}

\end{document}